\documentclass[prb,cha,nobibnotes,onecolumn,superscriptaddress,preprintnumbers,showpacs, amsmath,amssymb]{revtex4-1}
\usepackage{bm}
\usepackage{hyperref}
\hypersetup{colorlinks=true,linkcolor=black,citecolor=black,urlcolor=black}
\usepackage{amsmath}
\usepackage{amssymb}
\usepackage{amsthm}
\usepackage{amsfonts}
\usepackage{enumerate}
\usepackage{latexsym}
\usepackage{ifpdf}
\usepackage{graphicx}
\usepackage{makeidx}
\usepackage{multirow}

\expandafter\ifx\csname package@font\endcsname\relax\else
\expandafter\expandafter
\expandafter\usepackage
\expandafter\expandafter
\expandafter{\csname package@font\endcsname}
\fi
\usepackage[section]{placeins}
\usepackage{times}
\usepackage[normalem]{ulem}

\begin{document}

\title{Momentum-resolved electronic structures and strong electronic correlations in graphene-like nitride superconductors}

\author{Jiachang Bi}
\thanks{J.B. and Y.L. contributed equally to this work}
\affiliation{Ningbo Institute of Materials Technology and Engineering, Chinese Academy of Sciences, Ningbo 315201, China}
\affiliation{Center of Materials Science and Optoelectronics Engineering, University of Chinese Academy of Sciences, Beijing 100049, China}

\author{Yu Lin}
\thanks{J.B. and Y.L. contributed equally to this work}
\affiliation{Ningbo Institute of Materials Technology and Engineering, Chinese Academy of Sciences, Ningbo 315201, China}
\affiliation{Yongjiang laboratory, Ningbo, Zhejiang 315202, China}

\author{Qinghua Zhang}
\affiliation{Beijing National Laboratory for Condensed Matter Physics, Institute of Physics, Chinese Academy of Sciences, Beijing 100190, China}

\author{Zhanfeng Liu}
\affiliation{National Synchrotron Radiation Laboratory, CAS Center for Excellence in Nanoscience, University of Science and Technology of China, Hefei 230029, China}

\author{Ziyun Zhang}
\affiliation{School of Physical Science and Technology, Shanghai Tech University, Shanghai 201210, China}

\author{Ruyi Zhang}
\affiliation{Ningbo Institute of Materials Technology and Engineering, Chinese Academy of Sciences, Ningbo 315201, China}

\author{Xiong Yao}
\affiliation{Ningbo Institute of Materials Technology and Engineering, Chinese Academy of Sciences, Ningbo 315201, China}

\author{Guoxin Chen}
\affiliation{Ningbo Institute of Materials Technology and Engineering, Chinese Academy of Sciences, Ningbo 315201, China}

\author{Haigang Liu}
\affiliation{Shanghai Synchrotron Radiation Facility (SSRF), Shanghai Advanced Research Institute, Chinese Academy of Sciences, Shanghai 201204, China}

\author{Yaobo Huang}
\affiliation{Shanghai Synchrotron Radiation Facility (SSRF), Shanghai Advanced Research Institute, Chinese Academy of Sciences, Shanghai 201204, China}

\author{Yuanhe Sun}
\affiliation{Shanghai Synchrotron Radiation Facility (SSRF), Shanghai Advanced Research Institute, Chinese Academy of Sciences, Shanghai 201204, China}
\affiliation{Shanghai Institute of Applied Physics, Chinese Academy of Sciences, Shanghai 201800, China}

\author{Hui Zhang}
\affiliation{Shanghai Synchrotron Radiation Facility (SSRF), Shanghai Advanced Research Institute, Chinese Academy of Sciences, Shanghai 201204, China}

\author{Zhe Sun}
\affiliation{National Synchrotron Radiation Laboratory, CAS Center for Excellence in Nanoscience, University of Science and Technology of China, Hefei 230029, China}

\author{Shaozhu Xiao}
\email{shaozhu-xiao@ylab.ac.cn}
\affiliation{Yongjiang laboratory, Ningbo, Zhejiang 315202, China}

\author{Yanwei Cao}
\email{ywcao@nimte.ac.cn}
\affiliation{Ningbo Institute of Materials Technology and Engineering, Chinese Academy of Sciences, Ningbo 315201, China}
\affiliation{Center of Materials Science and Optoelectronics Engineering, University of Chinese Academy of Sciences, Beijing 100049, China}

\begin{abstract}

\textbf{ABSTRACT:}  Although transition-metal nitrides have been widely applied for several decades, experimental investigations of their high-resolution electronic band structures are rare due to the lack of high-quality single-crystalline samples. Here, we report on the first momentum-resolved electronic band structures of titanium nitride (TiN) films, a remarkable nitride superconductor. The measurements of crystal structures and electrical transport properties confirmed the high quality of these films. More importantly, with a combination of high-resolution angle-resolved photoelectron spectroscopy and the first-principles calculations, the extracted Coulomb interaction strength of TiN films can be as large as 8.5 eV, whereas resonant photoemission spectroscopy yields a value of 6.26 eV. These large values of Coulomb interaction strength indicate that superconducting TiN is a strongly correlated system.  Our results uncover the unexpected electronic correlations in transition-metal nitrides, potentially providing a perspective not only to understand their emergent quantum states but also to develop their applications in quantum devices.

\end{abstract}

\maketitle

\newpage

Electronic correlations, mainly arising from the Coulomb interactions among electrons, have been widely investigated in various types of materials such as transition-metal compounds (oxides and dichalcogenides), graphene, and metals. \cite{RMP-1998-Imada,PhysToday-2003-Tokura,Science-2000-Tokura, PRX-2017-Bisti,NC-2021-Whitcher, NatPhys-2019-Choi, Npj-2022-Mandal,NC-2022-Zhang, PRB-1984-Jensen,PRL-1989-Mahan} Particularly, electronic correlations play a central role in the emergence of unconventional superconductivity, colossal magnetoresistance, and charge (or orbital)  orders in transition-metal oxides. \cite{RMP-1998-Imada,PhysToday-2003-Tokura, Science-2000-Tokura} Interestingly, unlike strong electronic correlations in well-known transition-metal oxides and dichalcogenides, it is revealed that short-range electronic correlation can even be present in nearly free electron metals. \cite{Npj-2022-Mandal, NC-2022-Zhang, PRB-1984-Jensen, PRL-1989-Mahan} However, the understanding of momentum-resolved electronic structures as well as electronic correlations in transition-metal nitrides, a type of refractory quantum material that exhibits remarkable physical and chemical properties, is scarce due to the lack of high-quality single-crystalline materials. \cite{PRB-1980-Johansson,SolidState-1980-Johansson, PRB-1981-Johansson} 

Titanium nitride (TiN) with the cubic rocksalt structure ($ Fm\bar{3}m $, see the insert in Figure 1a) is a typical transition-metal nitride material that has been extensively applied in hard coatings, semiconductor chips, refractory plasmonics, and superconducting quantum devices. \cite{TheSolidFilms-1985-Sundgren,IEEE-2004-Chau, Science-2014-Guler,Science-2015-Boltasseva, JAP-2020-Richardson,APL-2013-Chang} More interestingly, several unexpected quantum phenomena have been observed in TiN, e.g., superconductor-insulator transition,\cite{PRL-2007-Baturina, PRL-2008-Sacepe} superinsulating states, \cite{Nature-2008-Vinokur} and the formation of cooper pairs above superconducting critical temperature. \cite{NC-2010-Sacepe,Science-2021-Bastiaans} Generally, these quantum phenomena in TiN can not be fully explained by a conventional physical picture. Therefore, the measurement of momentum-resolved electronic band structure and the understanding of electronic correlations in TiN are desirable at this moment. \cite{PRB-2009-Allmaier}

To address the above challenges, we synthesized a series of single-crystalline TiN films, the high qualities of which were confirmed by scanning transmission electron microscopy (STEM), high-resolution x-ray diffraction (XRD), low-energy electron diffraction (LEED), and electrical transport measurements. High-resolution electronic band structures of TiN films were uncovered by angle-resolved photoelectron spectroscopy (ARPES).  Combining ARPES spectra and the first-principles calculations, the extracted Coulomb interaction strength (\textit{U}) can be large as 8.5 eV, whereas resonant photoemission spectroscopy (resonant PES) yields the value of 6.26 eV.
These large values of Coulomb interaction strength indicate the strong electronic correlations in TiN films. 

\begin{figure*}[]
        \includegraphics[width=0.8\textwidth]{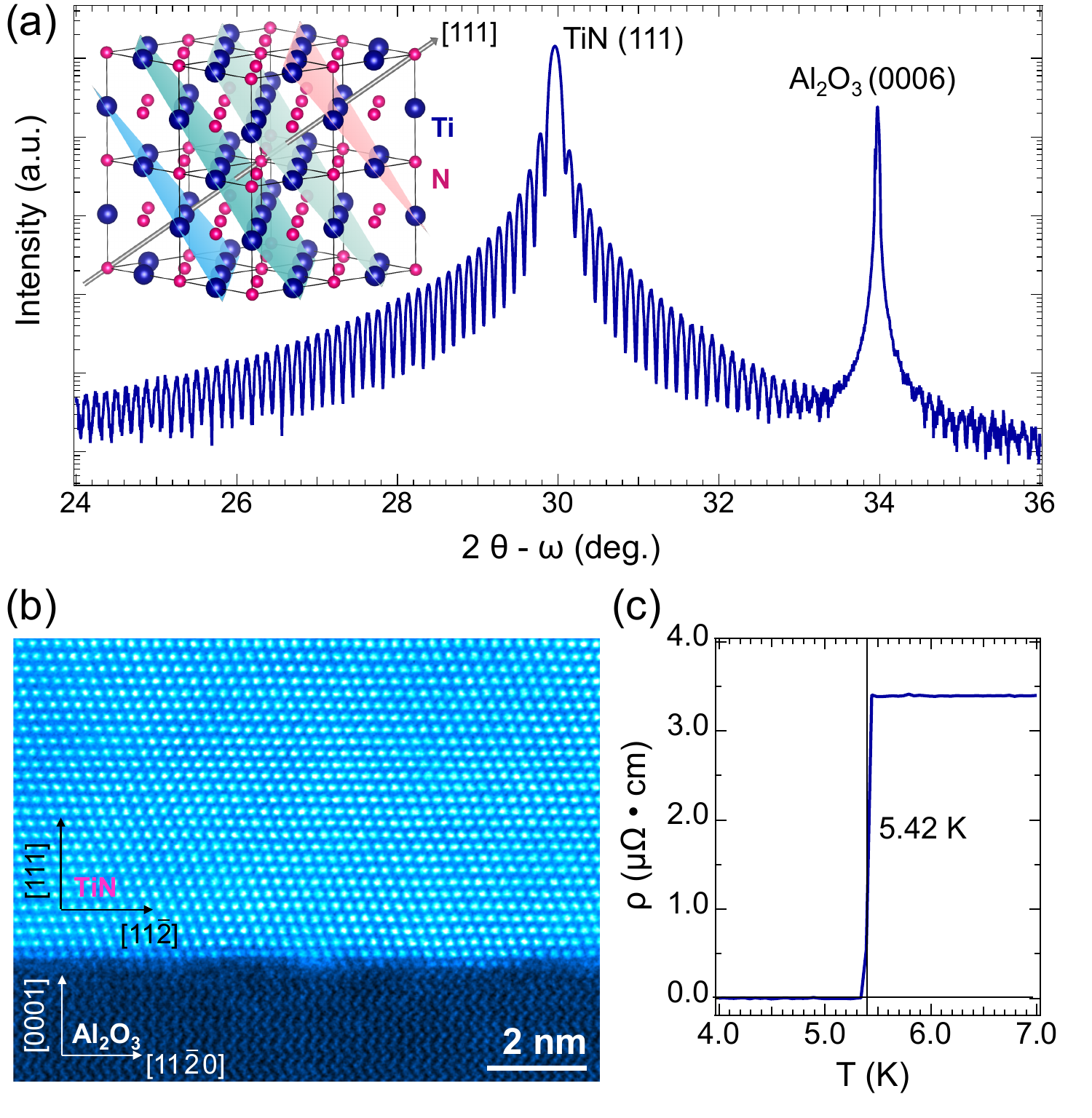}
        \caption{\label{} (a) The 2$ \theta -\omega$ scan around (111) diffraction of TiN films. The insert is the schematic of the TiN crystal structure with a rock-salt symmetry. The (111) planes are highlighted by colored planes. (b) The HAADF-STEM image of TiN films at the atomic scale. (c) Temperature-dependent resistivity of TiN films near the superconducting transition temperature (T$\rm_c$ $\sim$ 5.42 K).}
\end{figure*}

Epitaxial TiN (111) thin films ($\sim$ 75 nm and $\sim$ 10 nm) were synthesized by a homemade high-pressure magnetron sputter on $\alpha$-Al$_{2}$O$_{3}$ (0001) single-crystal substrates (5$\times$5$\times$0.5 mm$ ^{3} $).\cite{PRM-2021-Bi}
The crystal structures of TiN films were characterized by the lab-XRD, synchrotron-XRD, and transmission electron microscope. The electrical properties were measured by the Physical Property Measurement System (PPMS). After growth, the samples were transferred into the \textit{ex-situ} ARPES chamber.  The electronic structures of TiN films were investigated by the lab-ARPES, synchrotron-ARPES, and resonant photoemission spectroscopy (resonant PES). The electronic structures of TiN were calculated using the \textsc{Quantum Espresso} code, \cite{JPCM-2009-Giannozzi,JPCM-2017-Giannozzi} and the (111) surface spectra of TiN were calculated using the \textsc{WannierTools} package. \cite{CPC-2018-Wu} See the Supplementary Information (SI) for more experimental and calculational details. 

\begin{figure*}[]
        \includegraphics[width=0.95\textwidth]{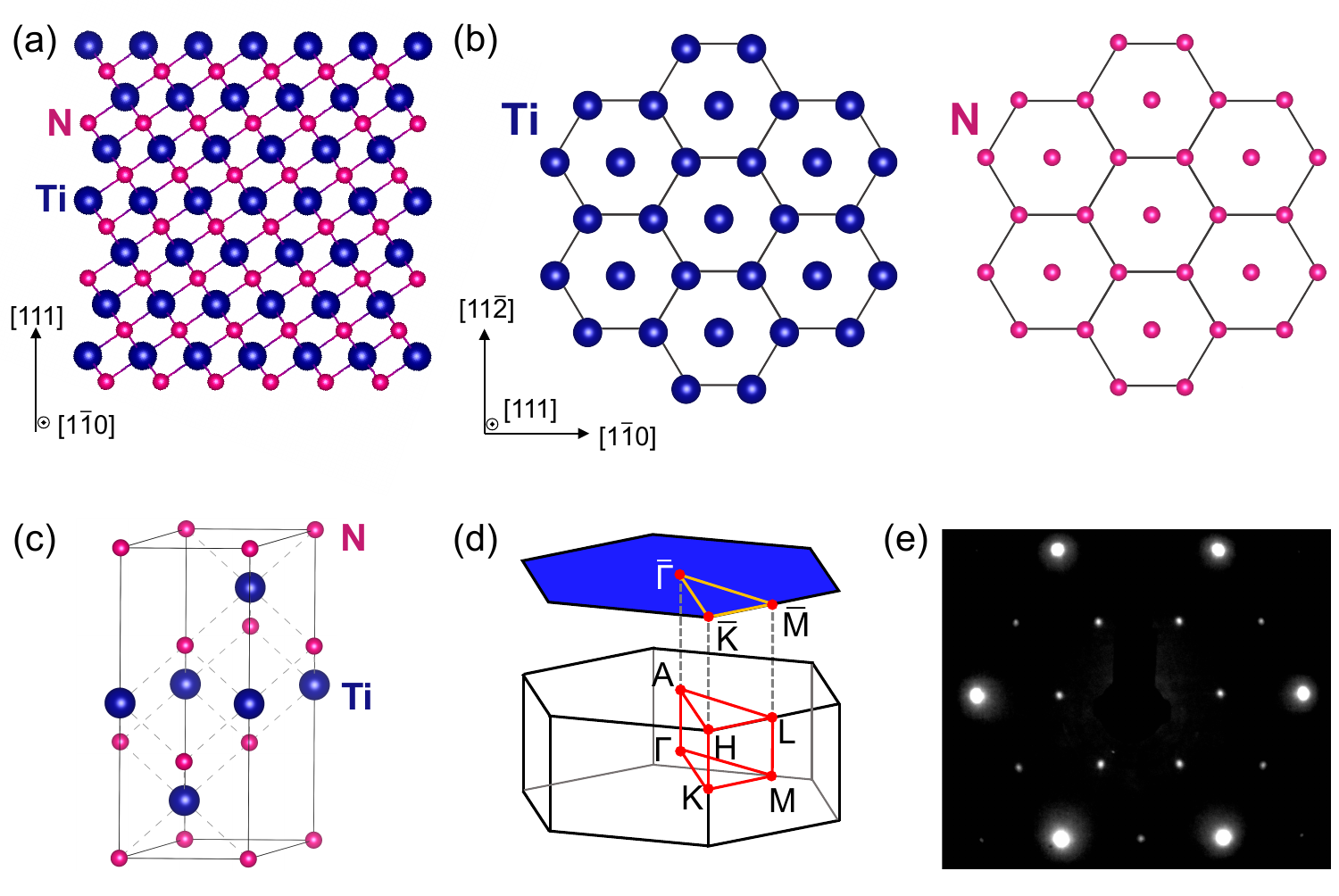}
        \caption{\label{} (a)  Side view of (111)-oriented TiN crystal structure, which is constituted of (b) alternate graphene-like Ti and N atomic layers. (c) The unit cell of TiN (111) for calculations.  (d) Three-dimensional Brillouin zone of TiN (111) and the projected (111) surface Brillouin zone. (e) LEED pattern of TiN (111) films with the electron energy at 110 eV.} 
\end{figure*}

First, we confirm the high quality of TiN films by high-resolution XRD, STEM, and electrical transport, which is a prerequisite for high-resolution ARPES measurements. The 2$ \theta-\omega $ scans of TiN films along the (111) reflection measured by synchrotron-XRD and lab-XRD were shown in Figure 1a and Figure S1b in SI, respectively. As can be seen, only the (111) diffraction peak is observable, and no secondary phase is detectable, indicating the epitaxial growth of TiN films. Moreover, the distinct thickness fringes (2$\theta$ between 24 to 36 degrees, see Figure 1a), the significant oscillations of x-ray reflection (XRR, see Figure S1a in SI), the narrow rocking curves (RC, see the insert in Figure S1b in SI), and the small surface roughness (see Figure S1e in SI) all confirm the high quality of the TiN films. Additionally, the (111) and (11$ \bar{2} $) layer spacing values extracted from reciprocal space mapping (RSM, see Figure S1d in SI) are around 2.452 and 1.728 \AA, respectively, which are close to the values of bulk TiN ($ d_{111} = 2.450 $ \AA, $ d_{11\bar{2}} = 1.732$ \AA), indicating the TiN film is almost fully relaxed. The atomic scale crystal structures of TiN films were characterized by high-resolution STEM. As shown in Figure 1b, the image of TiN film appears brighter than the Al$_{2}$O$_{3}$ substrate due to the scattering intensities being approximately proportional to the square of the atomic number (Z)\cite{Book-2017-Zuo}, and the bright feature mainly results from the Ti atoms. Whereas the signals of Al and O are weak (Al$_{2}$O$_{3}$ region) because of their small Z numbers. It is noteworthy that the TiN films are highly single-crystalline, and the interface between the TiN film and the substrate is atomically sharp. The temperature- and magnetic field-dependent resistivity data were shown in Figure 1c and Figure 2a,b in SI, from which the quality of TiN film can be further evidenced by the sharp transition and the high superconducting critical temperature (T$\rm_c$) near 5.42 K. It is noted that the T$\rm_c$ of TiN films grown by plasma enhanced molecular beam epitaxy (MBE) is around 5.2-5.4 K. \cite{JAP-2020-Richardson,JAP-2012-Krockenberger}

\begin{figure*}[]
	\includegraphics[width=0.95\textwidth]{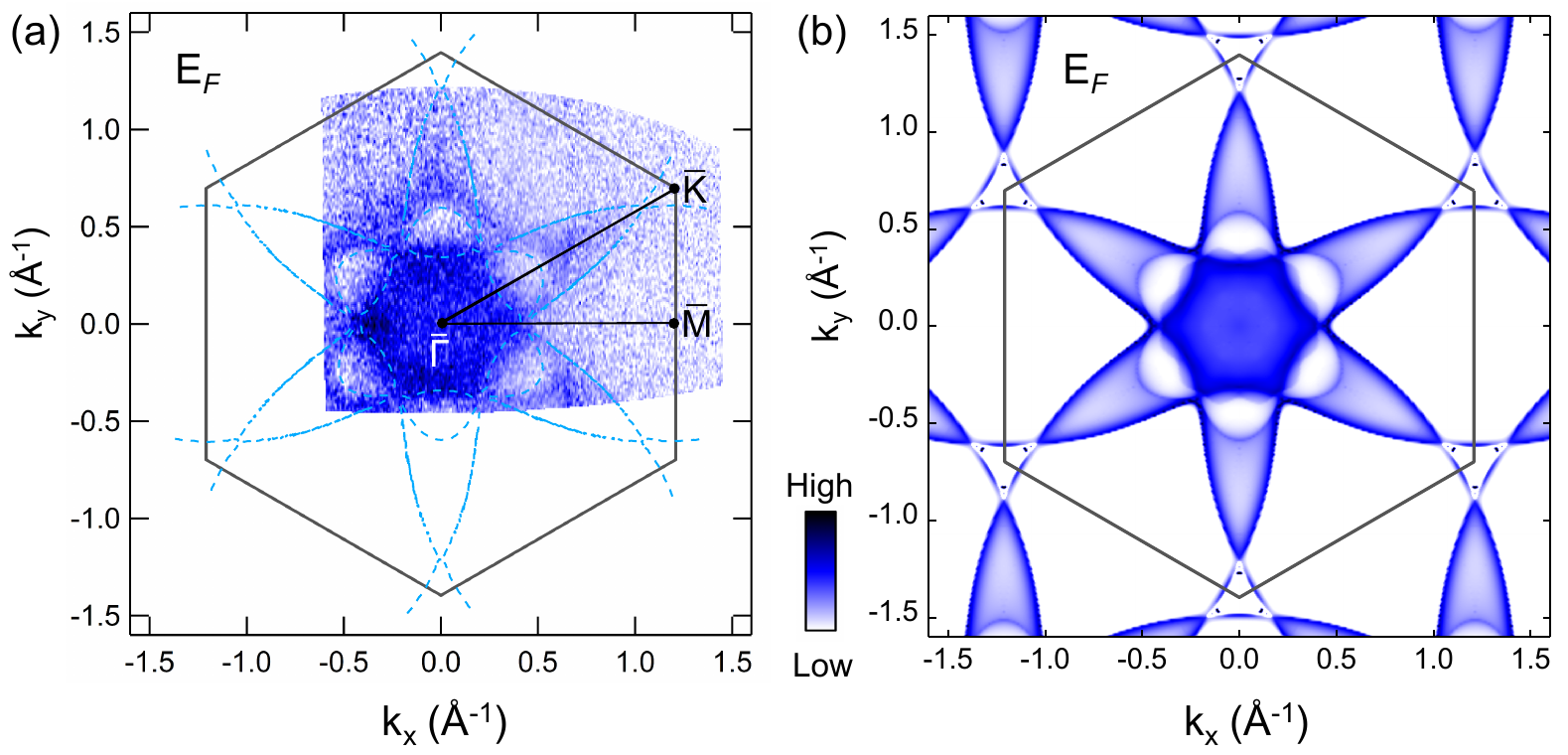}
	\caption{\label{}   (a) Experimental and (b) calculated (with $U$ $\sim$ 8.5 eV) Fermi surface mappings of TiN (111). The blue dotted curves in (a) are the outlines of calculated Fermi surface mappings from (b).}	
\end{figure*}

Next, we investigate momentum-resolved electronic structures of TiN (111) films by ARPES. As shown in Figure 2a,b, the crystal structure of TiN (111) is unique, which consists of alternate graphene-like Ti and N atomic layers. Figure 2c shows the unit cell of TiN (111) for the calculation. The three-dimensional (3D) bulk Brillouin zone (BZ) and the projected surface BZ (the blue plane) of TiN (111) can be seen in Figure 2d. Due to ARPES being a surface-sensitive probe, except the synchrotron-based x-ray ARPES, it is usually extremely challenging to probe the electronic structures of \textit{ex-situ} samples that have been exposed to the atmosphere. \cite{NanoLett-2019-Arab,PRL-2016-Vecchio,SciAdv-2021-Yu} Here, to remove the surface contaminations, TiN films were annealed at $\sim $500$^\circ$C for 7 hours before ARPES measurements. It is noteworthy that the effect of vacuum annealing on the superconducting transition temperatures of TiN films can be ignored (see Figure S2 in the SI). After annealing in the vacuum, a sharp LEED pattern (surface sensitive) with the expected hexagonal symmetry can be observed (see Figure 2e), further confirming the crystallinity of TiN films. Figure 3a shows the hexagonal petal-like Fermi surface (FS) of TiN (111) films, which is consistent with the six-fold symmetry of the TiN (111) in Figure 2e and Figure S1c in SI. Combined with the constant-energy contours at -0.5 eV (see Figure S3a in SI), it is revealed that the pockets around the $ \bar{\Gamma}$ point shrink with the energy away from the FS, indicating the type of pockets is electron-like. It is noted that both the FS (see Figure 3a) and constant-energy contours (see Figure S3a in SI) can be well reproduced by the calculations (see Figure 3b and Figure S3b in SI).
 \begin{figure*}[]
	\includegraphics[width=0.95\textwidth]{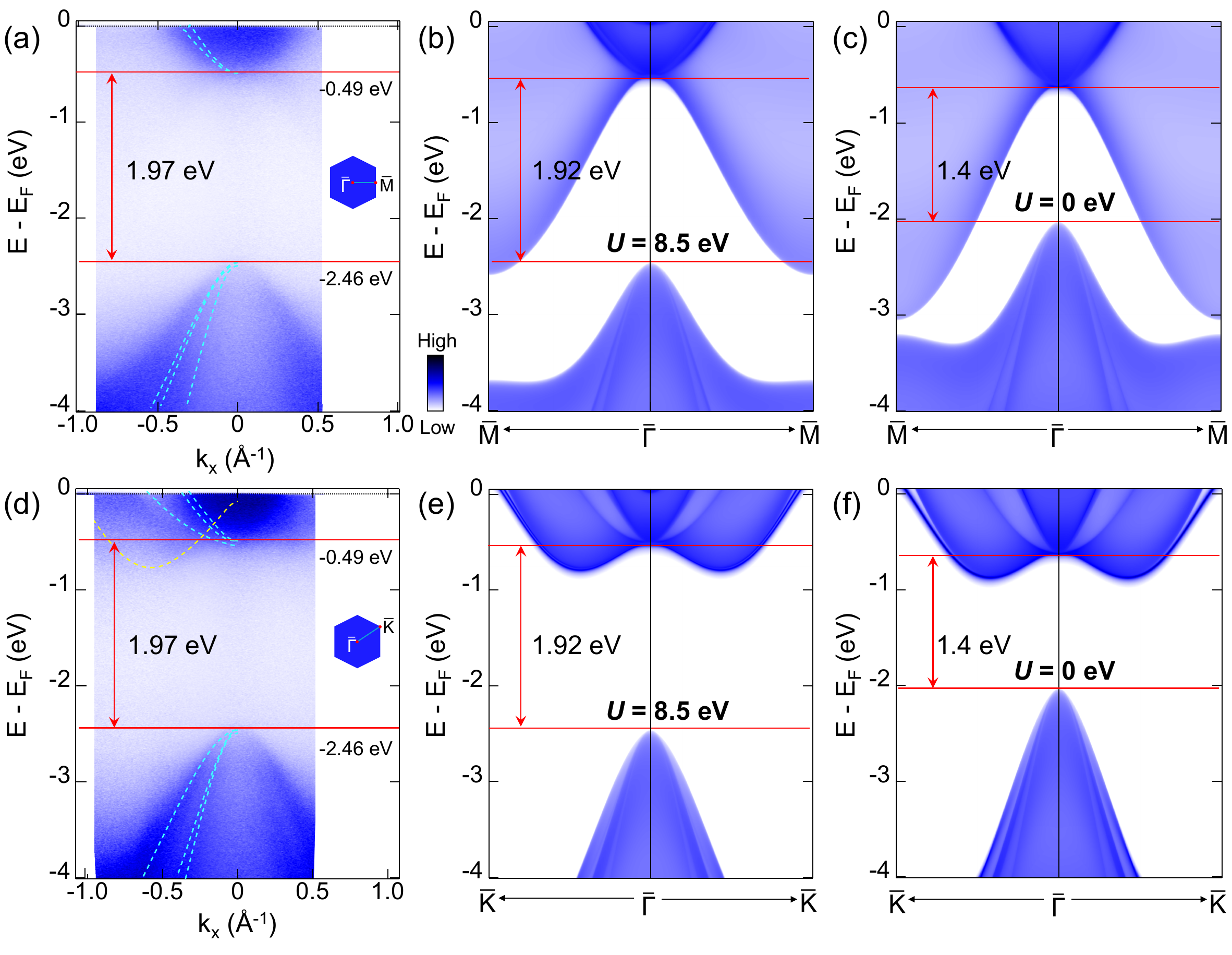}
	\caption{\label{} (a) Experimental band structures of TiN films along $ \bar{M}-\bar{\Gamma }-\bar{M}$. The red lines highlight the bottom of the conduction band (-0.49 eV) and the top of the valence band (-2.46 eV), yielding a band gap near 1.97 eV. The cyan curves are the calculated bulk bands from  $k_{z}$ = 0 plane (along ${M}$-${\Gamma }$) with $U$ = 8.5 eV. Calculated surface spectra of TiN (111) surface along $\bar{M}-\bar{\Gamma}-\bar{M}$ direction with $U$ = 8.5 eV (b) and without $U$ (c). The calculated band gaps are 1.92 and 1.4 eV, respectively, with and without $U$. (d) Experimental band structures of TiN films along $\bar{K}-\bar{\Gamma}-\bar{K}$ direction. The cyan curves are the calculated bulk bands from  $k_{z} = 0$ plane  (along ${K}$-${\Gamma }$), and the yellow ones are from  $k_{z} = \pi$ plane (along ${H}$-${A}$) with $U$ = 8.5 eV.  Calculated surface spectra of TiN (111) surface  along $\bar{K}-\bar{\Gamma}-\bar{K}$ direction with $U$ = 8.5 eV (e) and without $U$ (f). }
\end{figure*}

Figure 4 shows experimental and calculated electronic band structures of TiN films along the high-symmetry directions of $\bar{M}$-$\bar{\Gamma}$-$\bar{M}$ (see Figure 4a-c) and $\bar{K}$-$\bar{\Gamma}$-$\bar{K}$ (see Figure 4d-e), respectively. The signal-to-noise of the electronic band structures measured by the synchrotron-ARPES is better than that of lab-ARPES (see Figure S4 in SI). The significant features are the parabolic conduction bands and the anti-parabolic valence bands centered around the $ \bar{\Gamma}$ points in Figure 4a,d. At the $ \bar{\Gamma}$ point, the bottom of the conduction band is near -0.49 eV (indicated by the red lines at the top), whereas the top of the valence band is near -2.46 eV (indicated by the red  lines at the bottom). Naturally, at the $\bar{\Gamma}$ point, there is a band gap $\sim$ 1.97 eV (see the red arrows in Figure 4a,d) between the bottom of the conduction band and the top of the valence band. Unexpectedly, the experimental spectra can be reproduced very well with the Hubbard interaction ($U$) as large as 8.5 eV (-0.54, -2.46, and 1.92 eV, see Figure 4b,e). With this large $U$, not only the calculated bands from $k_z = 0$ plane (the cyan dotted curves in Figure 4a) along $\bar{M}$-${\Gamma}$-$\bar{M}$, but also the calculated bands from $k_z = 0$ plane (the cyan dotted curves in Figure 4d) or from $k_{z} = \pi$ plane (the yellow dotted curves in Figure 4d) along the $\bar{K}$-$\bar{\Gamma}$-$\bar{K}$  can agree well with the experimental spectra.  Additionally, the electronic band structures of thin TiN films ($\sim$ 10 nm) are analogous to thick TiN films ($\sim$ 75 nm), as seen in Figure S5 in SI. It is worth noting that the ARPES spectra usually involve the band contribution from nearly the entire $k_z$ range, which is a common phenomenon in low-photon-energy ARPES measurements due to the $k_z$ broadening effect. \cite{PRB-2018-Wang, PRB-2021-Xiao}  

However, without $U$, the calculated bottom of the conduction band at $\bar{\Gamma}$ point is near -0.63 eV, whereas the calculated top of the valence band is around -2.03 eV, yielding a band gap around 1.4 eV (see Figure 4c,f). As seen, these calculated band values (-0.63, -2.03, and 1.4 eV) without $U$ are in sharp contrast to the experimental values (-0.49, -2.46, and 1.97 eV). To further understand the role of $U$ in determining the electronic band structures, we plot the evolution of electronic band structures with variable \textit{U} (see Figure S6 in SI, \textit{U} = 0, 1, 3, 5, 6.5, 7.5, 8.5, and 9.5 eV). As seen, the effect of Hubbard interaction mainly changes the top of the valence bands, while the bottom of the conduction bands is almost unchanged. Thus, the band gap can increase from 1.4 eV to 1.92 eV when \textit{U} increases from 0 to 8.5 eV. Therefore, the combination of ARPES spectra and the first principles calculations demonstrate the strength of Coulomb interaction $U$ is as large as 8.5 eV, indicating the strong electronic correlations of TiN films. \cite{PRB-2009-Allmaier,PRB-1991-Anisimov,PRX-2017-Bisti,Npj-2022-Mandal}

\begin{figure*}[htbp]
	\includegraphics[width=0.95\textwidth]{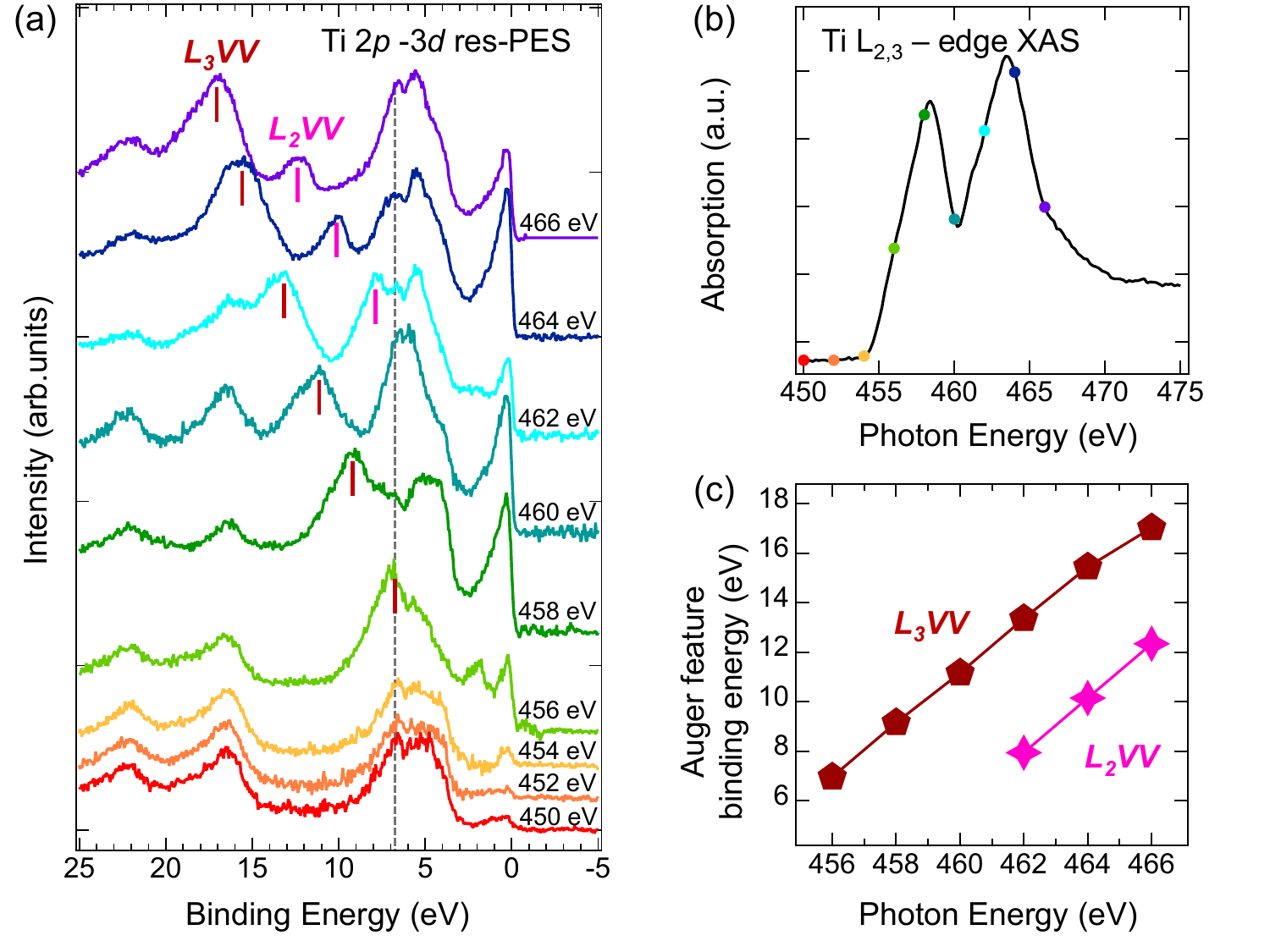}
	\caption{\label{}  (a) Resonant PES spectra of TiN films measured across the Ti 2$p$ -3$d$ threshold using incident photon energies indicated by the colored circles on the Ti L$_{2,3}$-edge XAS curve in (b). (c)  The plot of the binding energy of Auger features Ti L$_3VV$ and L$_2VV$ peaks as a function of the photon energy. }	
\end{figure*}

Finally, we estimate the value of $U$ experimentally. It is noteworthy that the values of $U$ in Cr metal,\cite{PRB-2000-Hufner} Fe metal,\cite{PRB-2000-Hufner} K$_2$Cr$_8$O$_{16}$,\cite{PRX-2015-Bhobe} TiSe$_2$,\cite{PRB-2020-Chuang} and CrN\cite{PRL-2010-Bhobe} compounds have been successfully extracted by the soft x-ray resonant photoemission (resonant PES). Figure 5a shows the incident photon energy ($h\upsilon$)-dependent resonant PES spectra of TiN film. As seen, the resonant PES spectra of TiN films are analogous when the incident photon energies (450 – 454 eV) are lower than the resonant energies (458.4 and 463.4 eV for L$_3$- and L$_2$-edges, respectively, see Figure 5b). However, with further increasing the incident photon energy from 456 to 466 eV, the two Auger peaks (Ti L$_3VV$ and L$_2VV$) become dominant and have strong dependences on the incident photon energy (see Figure 5a,c). The values of $U$ (6.26 $\pm$ 0.04 eV, see SI for details) for TiN films can be extracted by the formula $U =\left\{  E_{2p}-\left(  {h\upsilon-E_{LVV}}\right)\right\}-2\varepsilon_{3d}$,\cite{PRB-2000-Hufner,PRX-2015-Bhobe,PRB-2020-Chuang,PRL-2010-Bhobe} where $E_{2p}$ is the binding energy (BE) of Ti 2$p_{3/2}$ electron (455 eV, see Figure S7 in SI), $h\upsilon$ is the incident photon energy varied from 456 to 466 eV, and $E_{LVV}$  is the BE of the corresponding L$_3VV$ Auger peaks. The parameter $2\varepsilon_{3d}$ is the average energy of two uncorrelated holes, which can be obtained from the self-convolution of the single-electron Ti 3$d$ state.
Due to only the binding energy of 3\textit{d} electrons below the Fermi level contributing to the L\textit{VV} states, it is reasonable to take the value of $\varepsilon_{3d}$ as 0 eV for TiN. The estimated experimental value of \textit{U} (6.26 $\pm$ 0.04 eV, see SI for details) is smaller than that extracted from DFT calculations (\textit{U} = 8.5 eV). However, these values are larger than superconducting cuprates (Bi$_2$Sr$_2$Ca$_{0.92}$Y$_{0.08}$Cu$_2$O$_{8+\delta}$, $\sim$2.7 eV)\cite{PRB-2017-Yang} and the parent compound of superconducting nickelates (PrNiO$_2$, $\sim$5 eV)\cite{Matter-2022-Chen}. The extremely large values of $U$ extracted from the res-PES spectra further indicate that the superconducting TiN is a strongly correlated system.

Unlike conventional insulators without obvious critical transition temperatures, superinsulators are dualities of superconductors, \cite{Nature-2008-Vinokur,NP-2013-Ovadia} the conductivity of which can suddenly drop to zero at a certain temperature and then become superinsulating. Generally, the superinsulating behavior can only be present in the system with ultrahigh Coulomb repulsion between Cooper pairs. \cite{NP-2013-Ovadia,npj-2020-Rachmilowitz} It is noteworthy that the Coulomb interaction in two-dimensional systems can be several hundred meV or more, which is an order of magnitude stronger than that in three-dimensional systems.\cite{Innov-2021-Zou} Therefore, the identification of a large Coulomb interaction in this work can help to explain the presence of superinsulating behavior in the 2-dimensional TiN system.

In conclusion, we investigated the first momentum-resolved electronic band structures of TiN films by combining high-resolution ARPES, the first-principles calculations, and resonant PES. It is revealed that the extracted Coulomb interaction strength can be very large (8.5 eV from DFT-ARPES and 6.26 eV from resonant PES), indicating superconducting TiN is a strongly correlated material. Our results uncover the unexpected electronic correlations of transition-metal nitrides and possibly provide a perspective to understand the emergent quantum states (such as superconductor-superinsulator transitions and pseudo-gap states) in these compounds. More importantly, this work has demonstrated that high-resolution momentum-resolved electronic structures can be collected on the \textit{ex-situ} films by ARPES, which has been a big challenge for most epitaxial films. The combination of its uniquely resilient surface and strong correlated electronic states makes TiN as an ideal superconducting layer for the construction of novel superconducting Josephson junctions, topological superconducting heterostructures, and superconductor/ferromagnet heterostructures. These advancements pave the way for applications in superconducting qubits, topological qubits, and superconducting spintronics\cite{CM-2021-Kim, RMP-2008-Nayak, NP-2015-Linder, NRP-2020-Paschen}.

\section*{ASSOCIATED CONTENT}

\textbf{Supporting Information}

The Supporting Information is available free of charge at

The details of XRD, AFM, electrical transport, XPS, resonant PES, ARPES, and calculations.

\section*{AUTHOR INFORMATION}

\textbf{Notes}

The authors declare no competing financial interest.

\section*{ACKNOWLEDGMENTS}

We acknowledge insightful discussions with Jiandong Guo, Erjia Guo, Yue Cao, Xun Jia, Yilin Wang, Hanghui Chen, and Liangfeng Huang. This work was supported by the National Key R\&D Program of China (Grant No. 2022YFA1403000), the National Natural Science Foundation of China (Grant Nos. U2032126, U2032207, 11874058, 11974364, and 12204495), the Pioneer Hundred Talents Program of the Chinese Academy of Sciences, the Zhejiang Provincial Natural Science Foundation of China under Grant No. LXR22E020001, the Beijing National Laboratory for Condensed Matter Physics, and the Ningbo Science and Technology Bureau (Grant No. 2022Z086). This work is partially supported by the Youth Program of the National Natural Science Foundation of China (Grant No. 12004399).  We acknowledge the experiments support from BL02B of SSRF (proposal No.2023-SSRF-JJ-503256).

\begin{figure*}[htbp]
	\includegraphics[width=1\textwidth]{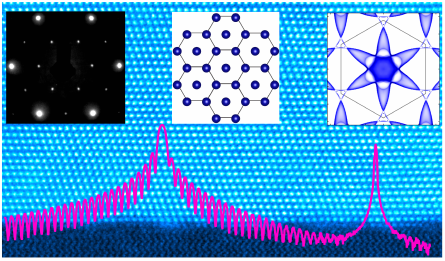}
\end{figure*}
\newpage

\end{document}